\documentstyle[]{article}
\begin{document}
 
\title{Rotation of the Universe and the angular momenta of celestial bodies}
\bigskip
 
\author{
W{\l}odzimierz God{\l}owski${^1}$
Marek Szyd{\l}owski ${^1}$ \\ Piotr Flin $^{2,3}$ 
Monika Biernacka${^2}$
}
 
\maketitle
 
1. Astronomical Observatory of the Jagiellonian University, 30-244
Krakow, ul. Orla 171, Poland  e-mail: godlows@oa.uj.edu.pl
 
2. Pedagogical University, Institute of Physics, 25-406 Kielce, ul.
Swietokrzyska 15, Poland  e-mail: sfflin@cyf-kr.edu.pl
 
3. Bogoliubov Laboratory of Theoretical Physics, Joint Institute for
Nuclear Research, Dubna, Moscow Region 141980, Russia
\bigskip

\section*{Abstract}
\medskip
We discuss the equation of motion of  the rotating homogenous and
isotropic model of the Universe. We show that the model
predicts the presence   of a minimum in the relation between the mass
of an astronomical object  and its angular momentum. We show that this
relation appears to be universal, and we predict the masses of
structures with minimal angular momenta in agreement with observations.
In such a manner we suggest the possibility at acquirement of angular momenta
of celestial bodies during their formation from the global rotation of the
Universe.

{\bf keywords}
 angular momenta, Universe rotation

\section{Introduction}

The pioneering idea of the rotation of the Universe
should be attributed to G. Gamow \cite{Gamow46},
who expressed the opinion that the rotation of galaxies is due
to the turbulent motion of masses  in the Universe,  and  ``we can ask
ourselves whether it is not possible to assume that {\it all matter in
the visible universe is in a status of general rotation around some
centre located far beyond the reach of our telescopes?}". The idea of
turbulence as a source of the rotation of galaxies was afterwards
developed by  C.F. von Weizsaeker \cite{Weizsaeker51},
 Ozernoy and Chernin \cite{Ozernoy68}, Ozernoy \cite{Ozernoy78}, but presently
is only of historical value.
[If the angular momenta of galaxies had orginated in such a way, their
spins should be perpendicular to the  main protostructure plane
\cite{Szandarin74}, which is not observed.]
The exact solution of the Einstein equation for the model of a homogeneous
universe with rotation and spatial expansion was proposed by Goedel
\cite{Goedel49, Goedel52}.
The observational evidence  of global rotation of the Universe was
presented by Birch \cite{Birch82}. He investigated the
position angles and polarisation  of classical high-luminosity  double
radio sources, finding that the difference between the position angle of
the source elongation and of the polarisation are correlated with the
source position in the sky. Immediately  there appeared
 a paper by Phinney and Webster \cite{Phinney83} concluding
that ``the data are unsufficient to substantiate the claim" and the statistics
are  applied incorrectly. Answering, Birch \cite{Birch83}
pointed out the difference in the quantity investigated by him and that by
Phinney and Webster, showing that their  data  exhibit the such effect.
 At the request ofBirch, Phinney and Webster \cite{Phinney84} reanalysed the
data, introducting new ``indirectional statistics" and taking into account
possible observational uncertainties. They  concluded ``that the
reported effect ({\it whatever may be its origin}) is strongly supported
by the observations".
Bietenholz and Kronberg \cite{Bietenholz84}
repeated the analysis for a larger sample of objects, finding no effect
of the Birch type.
New statistical tests were later applied to the data  \cite{Bietenholz86}.
 
Nodland and Ralston \cite{Nodland97a}
studied the correlation between the direction and distance to a
galaxy and the angle $\beta$ between the polarisation direction and the
major galaxy axis. They found  an effect of a systematic rotation of the
plane of polarisation of electromagnetic radiation, which  depends on redshifts.
As usually, the result was attacked for the point of incorrectly applied
statistics
\cite{Carrol97, Loredo97, Eisenstein97}
see the reply \cite{Nodland97b}
with a claim that the new, better data do not support the
existence of the effect \cite{Wardle97}.

The problem of the rotation of the whole Universe  has attracted the attention
of several scientists.
It was shown that the reported rotation values are too big when compared with
the CMB anisotropy.
Silk \cite{Silk70} pointed out that the dynamical effects
 of a general rotation of the Universe are presently unimportant, contrary
to the the early Universe, when  angular velocity $\Omega \ge 10^{-13}
rad/yr$.  He stressed that now the period of rotation must be greater than
the Hubble time, which is a simple consequence of the CMB isotropy.
Barrow, Juszkiewicz and Sonoda \cite{Barrow85} also addressed
this question. They showed that cosmic vorticity depends strongly on
 the cosmological models and assumptions connected with linearisation
of homogeneous, anistropic cosmological models over the isotropic
Friedmann Universe. For the flat universe, the value is ${\omega\over H_0}
\sim 2 \cdot 10^{-5}$.

Another interesting problem was the discussions on the empirical relation
between the angular momentum and mass of celestial bodies $J\sim M^{5/3}$
\cite{Brosche86}.
 Li-Xin Li \cite{Li98} explained this relation for galaxies as a
 result of the influence of the global rotation of the Universe on
galaxy  formation.

\section{ Universe and its angular momentum.}

Homogeneous and isotropic models of the Universe with matter may not
only expand, but also rotate relative to the local gyroscope. The
motion of the matter can be described by Raychaudhuri equation. This is
a relation between the scalar expansion $\Theta$, the rotation tensor
$\omega_{ab}$ and the shear tensor $\sigma_{ab}$ \cite{Hawking69} \cite{Ellis73}.
The perfect fluid has the stress-energy tensor:
$T_{ab}=(\rho+p)u_au_b+pg_{ab}$,
where $\rho$ is mass density and $p$ is pressure. The Raychaudhuri
equation can be written as:
    $$
-\nabla_a A^a+\dot{\Theta}+{1\over3}\Theta^2+2(\sigma^2-\omega^2)=
-4\pi G(\rho+3p),  \eqno(1)
    $$
where
$A^a=u^b\nabla_bu^a$  is the acceleration vector (vanishing in our case), while
$\omega^2\equiv\omega_{ab}\omega^{ab}/2$ and
$\sigma^2\equiv\sigma^{ab}\sigma_{ab}/2$ are rotation and sheer scalar
respectively, $\Theta$ is scalar expansion.

It has been shown that the spatial homogeneous, rotating  and  expanding
universe filled with perfect fluid must have non-vanishing  shear
\cite{King73}.
 
Because $\sigma$ falls off  more  rapidly  than  the  rotation $\omega$  as
the universes expand it is reasonable  to  consider  such  generalization  of
Friedmann equation in which only the ``centrifugal" term is present i.e.
      $$
{\dot{a} ^2\over 2} +{\omega^2a^2\over 2} -{4\pi Ga^2\over 3c^2}\epsilon=
 - {kc^2\over 2},  \eqno(2)
      $$
where $\epsilon=\rho c^2$ is energy density, $k$ is curvature constant,
 $a$ is scalar factor and $\dot{a} \equiv {d\over dt} a$
(or $^{\dot{}}\equiv {d\over dt}$ ).
Equation (2) should be completed with the principle of  conservation  energy
momentum (tensor) and that of angular momentum:
         $$
\dot{\epsilon}=-(\epsilon +p)\Theta, \qquad  \Theta\equiv 3{\dot{a} \over a} \eqno(3)
         $$
         $$
{p+\epsilon \over c^2} a^5 \omega =J.      \eqno(4)
         $$
From that we can observe that if  $p=0$  (dust)  then  $\rho\propto a^{-3}$
and $\omega\propto a^{-2}$, while in general $\sigma$ falls as $a^{-3}$
\cite{Hawking69}.
 The momentum conservation law should
be satisfed for  each kind of matter, and consequently the angular velocity of
the universe will evolve according to different  laws  in  different epochs.
Before
decoupling ($z=1000$), matter and radiation interact  but  after  decoupling
dust  and  radiation  evolve  separately  with their own   angular
velocities $\omega_d$ and $\omega_r$. Quantities $\omega$ and $\rho$ can be
written as $\omega=\omega_0 (1+z)^2$,
$\rho=\rho_{d0} (1+z)^3 +\rho_{r0}(1+z)^4$
the total mass density of matter and radiation.

 The conservation of the angular momentum of a galaxy
 relative to the gyroscopic frames in dust epochs gives \cite{Li98}:
     $$
J=kM^{5/3} -lM,                            \eqno(5)
     $$
where $k={2\over5}\left({3\over 4\pi\rho_{d0}}\right)^{2/3}\omega_0$,
$\rho_{do}$ is the density of matter in the present epoch,
$l=\beta r_f^2(1+z_f)^2\omega_0$, $r_f$ is galaxy radius, and
$\beta$ is a parameter determined by the distribution of mass in
the galaxy.
 
In \cite{Li98} the (present) value of the angular velocity of the Universe
is estimated. A suitable value for $k$ is 0.4 (in CGS Units). Taking
$\rho_{d0}= 1.88\cdot 10^{-29} \Omega h^2g\, cm^{-3}$
and $h=0.75$, $\Omega=0.01$
(Peebles \cite{Peebles93} for rich clusters of galaxies, see also
\cite{Peebles02, Lahav02}), we obtain
$\omega_0\simeq 6\cdot 10^{-21}rad\, s^{-1} \simeq 2\cdot 10^{-13}rad\, yr^{-1}$
 
It is interesting to note that there are the minimal values of $J_{min}$,
corresponding to same  $M_{min}$.
 
From the analysis of relation $J(M)$ [eq(5)], we obtain the presence of the
global minimum at
      $$
M_{min}=\left({3l\over 5k}\right)^{3/2}=1.95 r_f^3 (1+z_f)^3 \rho_{d0}, \eqno(6a)
      $$
      $$
J_{min}=-{6\sqrt{3}\over 25\sqrt{5}} {l^{5/2}\over k^{3/2}}, \eqno(6b)
      $$

For us it is important that $J$  grows as a  function of $M$  after the
minimal value of $M$.
It should be stressed that the value of $M_{min}$ {\it does not depend} on the
 value of $\omega$, i.e. the value of the rotation of the Universe.

Li-Xin Li \cite{Li98} considered the way  an object of the size of our
Galaxy is gaining angular momentum. It is an interesting approach to
the cosmogonical problem. Following  the considerations
 of Li-Xin Li \cite{Li98} by accepting $\Omega_m=0.01$, $z_f$
between 1 and 3, $r_f=30Kpc \approx 10^{23}cm $, and assuming the value of
$\beta$ $0.5$ or $0.4$ as the coefficient of the inertia momenta in the
equation for a celestial object (i.e assuming  disk like spherical shapes)
we obtain the value of $M_{min} \sim 5 \cdot 10^{39} g \sim 2.5\cdot 10^{3}
{\cal M}_{\odot}$. Fig. 1 shows dependence of $J(M)$ in that case.
 
\begin{figure}
\vskip 6cm
\includegraphics{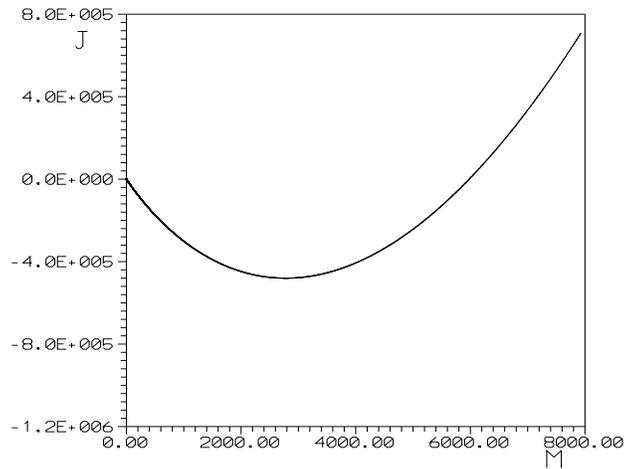}
\caption{The relation between angular momentum $J$ (in CGS units devided
by $10^{60}$) and M (in ${\cal M}_{\odot}$) of the astronomical object for
the galaxy-like protostructure.}
\end{figure}

From the observational point of view, only absolute values of $J$ in
relation (5) are important. Due to this fact, the minimum value of $|J|$
is easily observed. From  Equation (5) and (6a) it is seen
that this value equals 0 for:
      $$
M_{0}=\left({l\over k}\right)^{3/2} \approx 2.15 M_{min}. \eqno(7)
      $$
In the considered case $M_0 \approx 5\cdot 10^3$ ${\cal M}_{\odot}$.
This is sub-globular cluster mass. It seems to be accepted that such
structures are not rotating.
 
Because $M_{min}$ as well as $M_0$ do not depend on $\omega$, it is possible
to consider relation (6a) as a universal one for any collapsing-dust
proto-structure.
 
Let us consider a proto-solar cloud with a diameter of about $1$ $ps$. Because
the formation time of the solar system is certainly shorter than that of the
galaxy formation,  equation (7) gives  $M_{0} \approx 10^{24} g$.
Such are the masses of giant satellites in the Solar System. Disregarding
the Moon,  their angular momenta are  smaller than those of planets and
asteroids \cite{Wesson80}.  Thus, the  mass
corresponding to the minimal
momentum of a celestial body shows correctly those structures which in
reality have the minimal value of angular momenta.
 
\begin{figure}
\vskip 5cm
\includegraphics{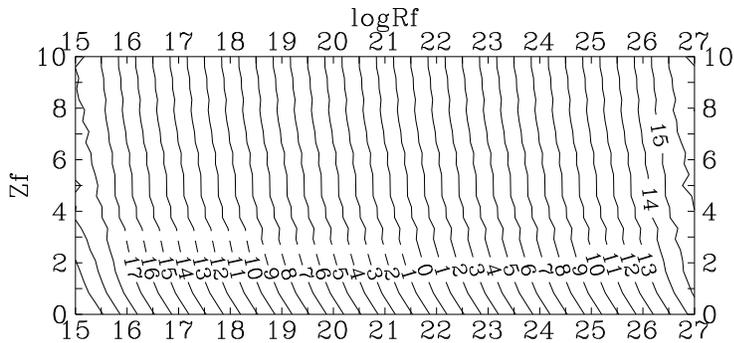}
\caption{The dependence beetween the value of $\log M_0$ (in ${\cal M}_\odot $),
logarithm the protostructure radius $r_f$ (in $cm$) and redshift formation $z_f$.}
\end{figure}
 
Numerical simulations with dark matter taken into account show that
primordial picture of large scale structure formation
consists of a network of filaments. During gravitational
collapse, clusters of galaxies  are formed at the intersections of filaments.
The question arises: how great $M_{0}$ (for dust) should be.
Assuming the radius of the proto-structure to be of the order of 30 Mpc,
which is consistent with the Perseus-Pisces and Hydra-Centaurus superclusters
\cite{Giovanelli88} and $z_f=6$  then we obtain
$M_0 \approx 5\cdot 10^{13} {\cal M}_{\odot}$. Taking into account that this is the
mass of dust, it corresponds to the total mass of galaxy cluster
 of the order of $10^{14}$ to $10^{15} {\cal M}_{\odot}$.
These contributions are consistent with observations under the assumption
that the evolution of dark matter density follows that of dust density.
We point out that presently there is no evidence of rotation of cluster of
galaxies.
In other words, our considerations show that the predicted masses of
structures having minimal angular momenta are in agreement with observations.
Assuming that the density, in which the proto-structure is formed is equal
to the dust density of the Universe, the radius of the proto-structure
together with the redshift formation univocaly determines the mass $M_0$
for which the absolute value of the angular momentum of the structure is
minimal. This relation is schematicaly presented in the Fig.2.
 In such a manner it is possible to consider a universal mechanism
of structure rotation.

{\bf Correspondence} should be addresed to W.Godlowski OA UJ Orla 171 Krakow
e-mail godlows@oa.uj.edu.pl fax 48-12-4251318


\begin{thebibliography}{}
\bibitem {Gamow46}
 Gamow, G. 1946 Nature 158, No 4016, 549 (1946)
\bibitem {Weizsaeker51}
 von Weizsaeker, C.F., ApJ 114, 165, (1951)
\bibitem {Ozernoy68}
 Ozernoy, L.M., Chernin, A.D., Astr. Zh., 45, 1137 (1968)
\bibitem {Ozernoy78}
 Ozernoy, L.M., 1978 in : Origin and Evolution of Galaxies and Stars (ed
Pikelner, S.B.,)  105, Nauka, Moscow (1978)
\bibitem {Szandarin74}
 Shandarin, S.F., Sov. Astr. 18, 392 (1974)
\bibitem {Goedel49}
 Goedel K., Rev. Mod.Phys 21, 447 (1949); GRG 32,1409 (2000),
\bibitem {Goedel52}
 Goedel K., In: Int. Cong. Math. ed L.M. Graves et al (1952); GRG 32, 1419 (2000).
\bibitem {Birch82}
 Birch P., Nature 298, 451 (1982)
\bibitem {Phinney83}
 Phinney E.S.,  Webster R.L., Nature 301, 735 (1983)
\bibitem {Birch83}
 Birch P., Nature 301, 736 (1983)
\bibitem {Phinney84}
 Phinney E.S., Webster R.L, Kendall D.G.,  Young G.A.,(1984) MNRAS 207, 637, (1984)
\bibitem {Bietenholz84}
 Bietenholz M.F., Kronberg,  ApJ 287, L1 (1984)
\bibitem {Bietenholz86}
 Bietenholz, M.F., AJ 91, 1249 (1986)
\bibitem {Nodland97a}
 Nodland, B., Ralston, J.P., Phys Rev Lett. 78, 3043 (1997)
\bibitem {Carrol97}
 Carrol, S.M., Field, G.B., Phys.Rev.Lett 79,2394, (1997)
\bibitem {Loredo97}
 Loredo, T.J., Flanganan, Wasserman, I.M., Phys.Rev.Lett D56, 7507 (1997)
\bibitem {Eisenstein97}
 Eisenstein, D.J., Bunn, E.F.,  Phys.Rev.Lett. 79, 1957 (1997)
\bibitem {Nodland97b}
 Nodlan, B., Ralston, J.P.,  Phys. Rev.Lett. 79, 1958 (1997)
\bibitem {Wardle97}
 Wardle J.F.C., Perley R.A., Cohen M.H., Phys. Rev. Lett. 79, 1801 (1997)
\bibitem {Silk70}
 Silk J., MNRAS 147, 13 (1970)
\bibitem {Barrow85}
 Barrow,J.D.,  Juszkiewicz, R., Sonoda, D.H., MNRAS 213, 917, (1985)
\bibitem {Brosche86}
 Brosche, P.,  Comm.Astroph. 11, 213 (1986).
\bibitem {Li98}
 Li-Xin Li., GRG 30, 497 (1998)
\bibitem {Hawking69}
 Hawking, S.W., MNRAS 142, 129 (1969)
\bibitem {Ellis73}
 Ellis, G.F.R., in Carges Lecture in Physics Vol 6 (ed. Schatzman E.) 1, New
 York Gordon and Brach Science Publishers (1973)
\bibitem {King73}
 King, A.R., Ellis G.F.R., 1973 Commun Math Phys 31 209 (1973)
\bibitem {Peebles93}
 Peebles, P.J.E., Principles of Physical Cosmology, Clarendon Press, Oxford (1993)
\bibitem {Peebles02}
 Peebles, P.J.E., Ratra, B., astro-ph/0207347 (2002)
\bibitem {Lahav02}
 Lahav, O., astro-ph/0208297 (1997)
\bibitem {Wesson80}
 Wesson, P.S.,  Astr. Astroph. 80, 296, (1980)
\bibitem {Giovanelli88}
 Giovanelli,R. Haynes M.P.  in Large Scale Motions in the Universe
 Princeton University Press, (eds V.C.Rubin G.V. Coyne), 31, Princeton (1988)
\end{thebibliography}
\end{document}